\documentclass[aps,twocolumn,nofootinbib]{revtex4-1}
\usepackage{amsmath}
\usepackage{amsfonts}
\usepackage{amssymb}
\usepackage{graphicx}
\usepackage{bm}
\usepackage{color}
\usepackage{ulem}
 
 \usepackage{epsfig}
\usepackage{graphicx}
\usepackage{chngcntr}
\counterwithout{figure}{section}
\usepackage{rotating}
\usepackage{amssymb}
\usepackage{amsmath}
\usepackage{physics}
\usepackage{multirow}
\usepackage{float}
\usepackage{caption, subcaption}

\begin{document} 

\title{\bf Emergence of a bicritical end point in the random crystal field Blume-Capel model}

\author{Sumedha,}
\email{sumedha@niser.ac.in}
\author{Soheli Mukherjee}
\email{soheli.mukherjee@niser.ac.in}
\affiliation{School of Physical Sciences, National Institute of Science Education and Research, Jatni - 752050,India}
\affiliation{Homi Bhabha National Institute, Training School Complex, Anushakti Nagar, Mumbai 400094, India}

\date{\today}

\begin{abstract}
We obtain the phase diagram for the Blume-Capel model with the bimodal distribution for random crystal fields, in the space of three fields: temperature($T$), crystal field($\Delta$) and magnetic field ($H$) on a fully connected graph. We find three different topologies for the phase diagram, depending on the strength of disorder. Three critical lines meet at a tricritical point only for weak disorder. As disorder strength increases there is no tricritical point in the phase diagram. We instead find a bicritical end point, where only two of the critical lines meet on a first order surface in the $H=0$ plane. For intermediate strengths of disorder, the phase diagram has critical end points along with the bicritical end point. One needs to look at the phase diagram in the space of three fields to identify various such multicritical points. 
\end{abstract}

\maketitle

\section{Introduction}

Multicritical points typically occur in systems described by three or more thermodynamic fields. In these systems, there can be critical points that can be reached only by fixing three or more thermodynamic parameters. Hence the full phase diagram of such systems is multi-dimensional \cite{knobler,fisher,griffithsh}. Such critical 
points are ubiquitous in nature, in systems like binary fluids \cite{widom,binarypd}, metamagnets \cite{metamagnet}, alloys of magnetic and non magnetic materials \cite{alloys}, $He^{3}-He^{4}$ mixtures \cite{helium}, quantum metals \cite{kirkpatrick}, polymer collapse \cite{degennes} and quantum chromodynamics \cite{qcd}. Among them the  tricritical point (TCP) is one of the most widely studied and well understood multicritical points \cite{lawrie}. Solvable models which display higher order critical points are useful in outlining the topology of the phase diagrams \cite{barbosa}. In this context, the mean field Blume-Capel model\cite{blume,capel} has been very useful and is 
one of the most well studied models. It is the simplest model to exhibit 
a TCP. TCP is an example of a multicritical point, which is a point of confluence of three critical lines in the space of three fields $(T,\Delta,H)$. Here $T$ and $H$ are the temperature and external field respectively and  $\Delta$ is a non-ordering field, known as the crystal field \cite{griffiths,lawrie}. In the $(T,\Delta)$ plane (with $H=0$), TCP shows itself as a point where the critical line ends in a first order line.

Introducing randomness in bond strength or field strength is known to affect the phase diagram. For example, in two dimensions it was shown that even an  infinitesimal amount of random field disorder can change a first order transition to a continuous transition or can destroy it altogether \cite{aizenman,huiberker}. In dimensions higher than two, similar behaviour is expected for strong disorder \cite{cardyj}. 
\begin{figure}[H]
\centering
\includegraphics[scale=0.42]{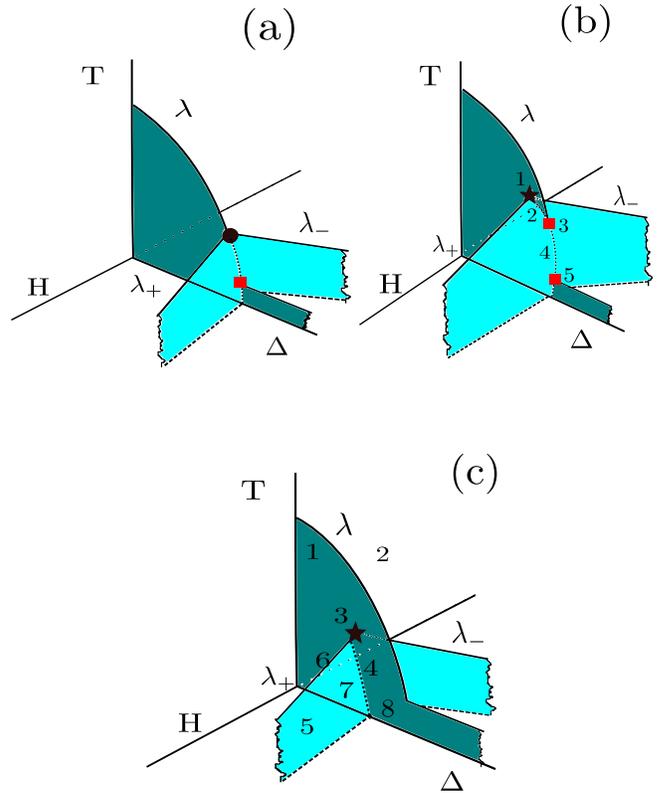}
\caption{Schematic phase diagram for different strengths of disorder: 
a) $0<p\le p_1$; b)$p_1<p\le p_2$ and c) $p_2 < p\le 0.5$. The value of $p$ represents the strength of disorder(there is no disorder for $p=0$ and the disorder is maximum for p=0.5). Solid lines represent lines of critical points and dotted lines represent first order transition lines. Solid dot represents TCP, solid square represents CEP and star represents BEP. Wiggled lines are to show the infinite length of wings. $\lambda$ represents the line of critical points in $H=0$ plane and $\lambda_+$ and $\lambda_-$ represent the critical lines for $H>0$ and $H<0$ respectively. The value of $p_1=0.022$ and $p_2=0.107875$ for the model studied in this paper}
\label{fig:1}
\end{figure}  

Blume-Capel model was introduced as an extension of Ising model where in addition to $\pm 1$, spin at each site can also be $0$. It has an extra term to take care of crystal field anisotropy \cite{blume,capel}. 
It was originally used to explain the first order magnetic transition in materials like $UO_2$ \cite{frazer}. Since then the model has found application in explaining the behaviour of wide range of physical systems like $He_3-He_4$ mixtures \cite{helium}, ternary fluids\cite{tfluids}, semiconductor alloys\cite{salloys}, phenomena of inverse melting\cite{imelting} and so on.  Study of these systems in random media is modelled by Blume Capel model with random crystal field \cite{buzano}. The phase diagram of the model is known to change under the effect of disorder. In particular, mean-field random crystal field Blume-Capel model has been studied extensively using many different techniques like mean field approximation \cite{mfdilution},effective field theories\cite{yuksel}, renormalisation group\cite{branco,snowman}, Bethe lattice\cite{bethe},pair approximation method \cite{lara},replica \cite{salmon,santos} and large deviations \cite{sumedha1}. All these work have focussed on the $(T,\Delta)$ plane. These different methods do not agree with each fully in the prediction of the phase-diagram, but they all report that the first order line and hence the TCP disappears for higher strengths of disorder. Some of them \cite{mfdilution,salmon, santos} predicted different topologies of the phase-diagram depending on the strength of disorder, with multicritical points like critical end points, ordered critical point and double critical point. 

Since TCP is a point of confluence of three critical lines, it is important to look at the effect of disorder on the other two lines meeting at the TCP. Hence, we revisit the problem and obtain the phase diagram in the space of three fields, on a fully connected graph by solving the model exactly, using large deviation theory \cite{touchette}. 
 For the  Blume Capel model, the mean field solution is known to give the correct values of the critical exponents at the TCP in three dimensions \cite{lawrie} and predicts the correct topology of the phase diagram for $d \ge 2$, in the pure case\cite{zirenberg,silva}. Similarly,  we expect the change in topology of the phase diagram as a function of disorder strength derived in this paper to be robust and not restricted to mean-field solutions.
Note that Landau approach cannot be used when the external field $H \neq 0$, as the value of the magnetisation is finite along the critical lines. We hence have to make use of the full free energy functional even to determine the critical lines.
 
We find that the TCP persists for only very weak disorder strengths. As the disorder strength increases, the TCP vanishes and a different multicritical point, bicritical end point(BEP) emerges where only two of the three critical lines end on a first order surface \cite{nelson,helena-barbosa,plascak}. This point has been wrongly reported as an ordered critical point in earlier studies in $(T,\Delta)$ plane \cite{santos}. BEP has been comparatively less observed and studied in the literature. Two well known examples where BEP has been observed are: anisoptropic continuous spin systems as an end point of the spin flop line\cite{nelson} and in spin 3/2 systems with crystal field\cite{plascak,butera}. We also find that the model exhibits critical end points(CEP) for intermediate strengths of disorder as reported in earlier studies \cite{santos}. Critical end point is a critical point where a line of second order transitions terminates at a line of first order transitions \cite{chaikin}. Alternately, it can also be defined as a point where two phases become critical in the presence of one or more ordered phases, known as the spectator phases \cite{upton}, in systems with multiple phases. We thus find three different phase diagrams depending on the strength of disorder. Recall that for the pure Blume Capel model, the phase diagram has three critical lines ($\lambda, \lambda_+,\lambda_-$) which all meet at the tricritical point \cite{lawrie}. In the mean field limit, all the critical points along these three lines fall in the Ising universality class. Along $\lambda$-line there is a spontaneous symmetry breaking transition from state with magnetisation $m=0$ to a state with $|m| \neq 0$ in $(T,\Delta)$ plane. Switching on the magnetic field introduces bias toward $m_+$ or $m_-$ state, depending on the sign of the magnetic field. This results in the coexistence of $m=0$ state with $m_+(m_-)$ states for low $H_+(H_-)$ respectively. These two coexistence surfaces meet along a triple line in the $(T,\Delta)$ plane. 
$\lambda_+ / \lambda_-$ lines separate this coexistence surface from 
the ferromagnetically ordered phases with opposite magnetisation. Hence the pure Blume Capel phase diagram has only one ferromagnetic state and one paramagnetic state in the $(T,\Delta)$ plane. 

We find that for weak disorder, the three critical lines meet at a tricritical point(see Fig1(a)). But a new ferromagnetic state appear now at very low temperature for all $p>0$, which is separated from usual ferromagnetic state via first order quadruple line, that ends in a CEP.
The value of magnetisation in this phase depends on the strength of disorder and increases with increasing disorder strength. 

 For intermediate disorder strengths, the two critical lines with $H \neq 0$ meet at a BEP and the line of continuous transition in $(T,\Delta)$ plane (known as $\lambda$ line) meets a line of first order transition at a CEP. CEP and BEP are connected via a quadruple line, along which the four phases co-exist(see Fig 1(b)). The quadruple line separates two ferromagnetic phases. This happens due to abrupt change in the number of $\pm 1$ spins across this line. Unlike the new ferromagnetic state that occurs at low temperature, entropy is important for this ferromagnetic state as it occurs at a relatively higher temperature and hence entropically it becomes useful to have more spin 
 particles. The system hence has three ferromagnetic and one paramagnetic phase in this range of disorder.
 
  For strong disorder, the BEP persists but CEP vanishes and the $\lambda$ line continues to $\Delta \rightarrow \infty$ (see Fig. 1(c)). This phase has the same two ferromagnetic phases as the weak disorder case and a paramagnetic phase. We will study these three topologies in this paper. 

The plan of the paper is as follows: In Section \ref{sec1} we discuss the Blume-Capel model in the presence of external field and derive the equations for critical lines in $(T,\Delta,H)$ space. In Section \ref{sec2} we study the phase diagram for strong disorder and intermediate disorder by using the full free energy functional. We also look at  the magnetisation, density and magnetic susceptibility near the BEP, to understand the nature of BEP. In Section \ref{sec3} we briefly discuss the case of weak disorder and in Section \ref{sec4} we show that a Landau expansion of the free energy functional cannot describe the BEPs and CEPs of this model. We conclude with a short discussion in Section \ref{sec5}.

\section{Model}\label{sec1}
We study the Blume-Capel model with random crystal field disorder in the presence of external field on a fully connected graph. The Hamiltonian can be written as 
\begin{equation}
H(C_N) = -\frac{1}{ 2N} (\sum_{i} s_i)^2 -\sum_{i} \Delta_{i} s_{i}^2 -H \sum_{i} s_i
\end{equation}
where $\Delta_i$ represent quenched random crystal field at each site, $H$ is the external field and $s_i$ are spin$-1$ random variables which can take $\pm 1,0$ values. There are two order parameters: magnetisation, $m = \bar{s}$ and density of $\pm 1$ spins, $q = \bar{s^2}$. These are obtained by taking a quenched average of the random variables $s$ and $s^2$ respectively \cite{cardy}. We draw random crystal fields from bimodal distribution of the kind:  
\begin{equation}
P(\Delta_i) = p \delta(\Delta_i -\Delta)+(1-p) \delta(\Delta_i+\Delta)
\end{equation}
Since $p=0$ or $1$ will imply no disorder and $p=1/2$ would be the most random case, it is enough to look for $0 \leq p  \leq 0.5$. 

It can be shown that the probability of a configuration $C_N$ satisfies large deviation principle(LDP) in the presence of random crystal field disorder\cite{touchette,ellis,sumedha1},i.e
\begin{equation}
P(C_N:\sum_i s_i =x_1 N;\sum_i s_i^2=x_2 N) \sim \exp(-N I(x_1,x_2))
\end{equation}

The rate function $I(x_1,x_2)$ for bimodal random crystal field disorder in the absence of external field was calculated using tilted LDP recently\cite{sumedha1}. Using the same method, the rate function in the presence of external field is:
\begin{eqnarray} \label{eq:5}
I(x_1,x_2) &=& x_1 \tanh^{-1} \Big(\frac{x_1}{x_2}\Big) + x_2 \Big[\ln \frac{z}{2 \cosh(\tanh^{-1}\frac{x_1}{x_2})}\Big] \nonumber \\
&&-p \ln (1+ z e^{\beta \bigtriangleup})-(1-p) \ln (1+ z e^{-\beta \bigtriangleup})\nonumber \\
&&+ p \ln (1+ 2 e^{\beta \bigtriangleup})+(1-p)\ln (1+ 2 e^{-\beta \bigtriangleup}) \nonumber \\
&&-\frac{\beta x_1^2}{2}-\beta H x_1
\end{eqnarray}
where $z$ is the solution of the equation:
\begin{equation}
\label{zeq1}
\frac{x_2}{z}=\frac{p e^{\beta \Delta}}{1+z e^{\beta \Delta}}+\frac{(1-p) e^{-\beta \Delta}}{1+z e^{-\beta \Delta}}
\end{equation}

In the limit, $N \rightarrow \infty$, for a given $\beta ,\Delta$ and $H$, the value of $x_1$ and $x_2$ that minimise $I(x_1,x_2)$ will give the value of magnetisation ($m$) and density($q$). The minima of the rate function in $(x_1,x_2)$ plane gives the free energy for a given $\beta(=1/T), \Delta$  and $H$. Hence the values of $x_1$ and $x_2$ which minimise $I(x_1,x_2)$ are the value of $m$ and $q$ respectively for a given set  of thermodynamic variables. Minimising $I(x_1,x_2)$ with respect to $x_1$ and $x_2$ results in the following equations for $m$ and $q$: 
\begin{equation}
\label{opeq1}
\tanh(\beta (m+H))=\frac{m}{q}
\end{equation}
\begin{equation}
\label{opeq2}
z=\frac{2}{\sqrt{1-m^2/q^2}}
\end{equation}
where $z$ is related to $q$ via the Eq. \ref{zeq1}, i.e:
\begin{equation}
\label{zeq}
\frac{q}{z}=\frac{p e^{\beta \Delta}}{1+z e^{\beta \Delta}}+\frac{(1-p) e^{-\beta \Delta}}{1+z e^{-\beta \Delta}}
\end{equation}

In subsection \ref{sec1a}  we will recap the results in the absence of external field and then build the equations for phase diagram in $(T,\Delta,H)$ space in subsection \ref{sec1b}

\subsection{Two field phase diagram in the $(T,\Delta)$ plane}
\label{sec1a}

For $H=0$, the phase diagram has been studied earlier \cite{sumedha1,santos}. We will briefly recap those results here: Assuming $m$ to be small the fixed point equations, Eq. \ref{opeq1} and \ref{opeq2} can be linearized around $m=0$. This gives $q=1/\beta$ and $z=2$ at the  critical point. Substituting these values in Eq. \ref{zeq}, gives the equation for a line of continuous transition in the $H=0$ plane. The line of continuous transition in $H=0$ plane is known as the $\lambda$-line and satisfies the following equation 
\begin{equation}\label{lline}
5-4 \beta = 2 (\beta p-1) e^{\beta \Delta}+2(\beta-\beta p-1) e^{-\beta \Delta}
\end{equation}

This is valid only when the higher order terms in the expansion can be ignored. Taking $q=(1+\epsilon)/\beta$ and expanding in powers of $\epsilon$ we find that  the coefficient of linear term in $\epsilon$ becomes zero when

\begin{equation}
\label{lap}
12 \beta -17+(3 \beta -10) \cosh(\beta \Delta)-3 \beta (1-2 p) \sinh(\beta \Delta) =0
\end{equation}

Solving Eq. \ref{lline} and \ref{lap} together we get the condition for break down of linear approximation as
\begin{equation}\label{ctcp}
cosh(\beta \Delta)= \frac{12 \beta-19}{8}
\end{equation}
Hence for a given $\Delta$, there will be either no transition or a first order transition, beyond the value of $\beta$ that satisfy Eq. \ref{ctcp}

The value of $(\beta,\Delta)$(or equivalently $(T,\Delta)$) which satisfy Eqs. \ref{lline} and Eq. \ref{ctcp} simultaneously gives the location of TCP for a given $p$. It was found in \cite{sumedha1} that beyond $p_c=0.0454$ the two equations cannot be satisfied simultaneously and hence there is no TCP, and the $\lambda$ line in the $(T,\Delta)$ plane extends to $\Delta \rightarrow \infty$. This treatment is equivalent to Taylor expanding the rate function to get an equivalent Landau free energy functional, which we will discuss in Sec. \ref{sec4}.

\subsection{Three field phase diagram in $(T,\Delta,H)$ space}
\label{sec1b}
Let us now take $H \neq 0$ and look for the critical points in the full ($T,\Delta,H$) space. We know that at the TCP there are two other continuous lines with $H \neq 0$ which meet the $\lambda$ line. We call these,  depending on the value of $H$, as $\lambda_+$ and $\lambda_-$. 

We wish to understand the effect of disorder on the two critical lines $\lambda_+$ and $\lambda_-$. We will focus on the effect of disorder on these two critical lines in this paper. Along these lines, $m \neq 0$ and one cannot look for continuous transition by expanding the free energy functional like we did in the section \ref{sec1a}. 

Note that at the fixed point the value of $m$ and $q$ are related via Eq. \ref{opeq1}. Since we are only interested in the fixed points, at fixed points $q$ is completely determined by $m$, hence the 
rate function can be replaced by a one parameter functional $\tilde{f}(m)$, which comes out to be:

\begin{eqnarray} \label{eq:4}
\tilde{f}(m) &= & \frac{\beta m^2}{2} -p \log(1+2 e^{\beta \bigtriangleup} \cosh \beta (m+ H)) \nonumber \\
&&- (1-p)\log (1+2 e^{-\beta \bigtriangleup} \cosh \beta(m+ H) )+ \nonumber \\
 && p\log (1+2 e^{\beta \bigtriangleup})+(1-p)\log (1+2 e^{-\beta \bigtriangleup})
\end{eqnarray}
From this  we get the following self-consistent equation for $m$:
\begin{eqnarray} \label{eq:m}
 m &=&2  \sinh \beta(m+H)  \Bigg [\frac{p e^{\beta \bigtriangleup}}{1+ 2 e^{\beta \bigtriangleup} \cosh\beta(m+H)}+ \nonumber \\&&\frac{(1-p) e^{-\beta \bigtriangleup}}{1+ 2 e^{-\beta \bigtriangleup} \cosh\beta(m+H)} \Bigg] 
\end{eqnarray}

Since $m \neq 0$ along the $\lambda_+$ and $\lambda_-$ lines, expanding  $\tilde{f}(m)$ in powers of $m$ to get a Landau free energy functional will not give the correct critical behaviour. But, in general along a critical line, the first three derivatives of the free energy functional with respect to the order parameter should be zero. This is because between two successive minimas , there must exist two points of inflexion, i.e $f^{''}=0$ and hence also a point where $f^{'''}=0$. Hence at the continuous transition, all three derivatives should vanish simultaneously. Hence to study $\lambda_+$ and $\lambda_-$ critical lines we equate the first three derivatives of $\tilde{f}(m)$ w.r.t $m$ to zero \cite{lawrie}(and fourth derivative should be greater than zero). This is true also for the $\lambda$ line, as for $H=0$ and $m=0$ the third derivative is trivially zero and second derivative gives the same condition as Eq. \ref{lline}. 

In general, equating second and third derivative of $\tilde{f}(m)$ to zero we get the following two conditions respectively:
\begin{eqnarray}
 && \frac{p(2 x^2+ x y)}{(1 +2 xy )^2}+\frac{(1-p)(2+ x y)}{(x +2y )^2} =\frac{1}{2 \beta} \label{eq:1} \\
&& \frac{p(x -8 x^3- 2 x^2 y)}{(1 +2xy )^3}+\frac{(1-p)(x^2 -8- 2 xy)}{(x +2y )^3} = 0\label{eq:2}
\end{eqnarray}
here $x=\exp(\beta \Delta)$ and $y=\cosh \beta(m+H)$. For $p \neq 0$, the two equations are quartic and hexic in $x$. 

For $p=0$, they reduce to the following simpler equations :
\begin{eqnarray}
&& \frac{2+ x y}{[x +2y ]^2} =\frac{1}{2 \beta} \label{purewings1}\\
&& \frac{x^2 -8- 2 xy}{[x +2y ]^3} =0
\label{purewings2}
\end{eqnarray}
Solving these equations we get
\begin{eqnarray}
y &= & \cosh\beta(m+H)=\frac{\beta -2}{\sqrt{4-\beta}} \\
x &=& e^{\beta \bigtriangleup}=\frac{4}{\sqrt{4-\beta}}
\end{eqnarray}
Hence, we reproduce the classic results of Blume,Emery and Griffiths \cite{beg}: There is a line of critical points for $4\geq \beta \geq 3$ for $H> 0$ and another for $H<0$. Both critical lines extend to $\Delta \rightarrow \infty$. These two lines enclose two first order surfaces which meet in the $H=0$ plane along a triple line(line with three phase co-existence). Above $\beta=4$ there is no value of $x$ and $y$ that can satisfy Eqs. \ref{purewings1} and \ref{purewings2} simultaneously. The magnetisation along these two critical lines is not zero and is equal to
\begin{eqnarray}
m &=&  \pm \sqrt{\frac{\beta -3}{\beta}}
\end{eqnarray}
This can be used to get the value of $H$ along the critical lines, which comes out to be
\begin{eqnarray}
H=\pm \frac{1}{\beta}\log(\frac{\beta-2+\sqrt{\beta^2-3\beta}}{\sqrt{4-\beta}})- m
\end{eqnarray}
These two critical lines meet in the $H=0$ plane at a point with $T_{TCP}=1/3$ and $\bigtriangleup_{TCP}=0.462098$. This is the well known TCP in ($T,\Delta$) plane for $p=0$(can be obtained by solving  Eq. \ref{lline} and \ref{ctcp} simultaneously for $p=0$).

For $p \neq 0$, we use Mathematica \cite{mathematica} to solve Eq(\ref{eq:1}) and Eq(\ref{eq:2}) simultaneously to  get the two critical lines numerically. To solve the equations for any arbitrary $p$, we scan different values of $\beta$ and $\Delta$  and hence $x$ and solve Eq(\ref{eq:1}) (corresponding to $\tilde{f}''(m)=0$) exactly to get the corresponding value of $y$. Then we substitute the value of $x$ and $y$ in Eq(\ref{eq:2}) to check if $(x,y)$ satisfy the condition, $\tilde{f}'''(m)=0$. 

For each set of $(x,y)$  that satisfy Eq. \ref{eq:1} and Eq. \ref{eq:2} simultaneously, we can calculate $m$ using the equation:
\begin{equation}
m =\pm 2  \sqrt{y^2 -1} \Big[\frac{p x}{1+ 2 x y}+\frac{(1-p) }{y+ 2 x}\Big]
\end{equation}
The above equation is derived from Eq. \ref{eq:m} by taking 
$\cosh \beta(m+H)=y$ and $\exp(\beta \Delta)=x$. The corresponding value of $H$ along the critical lines can then be calculated by inverting $y=\cosh \beta(m+H)$.

For a TCP to exist the two critical lines in the $H \neq 0$ plane should meet in $H=0$ plane at the point where second order line ends in a first order transition line in the ($T,\Delta$) plane. We can put $H=0$ and $m=0$ in Eqs.  \ref{eq:1} and \ref{eq:2} to directly look for this point. Hence, we separately solve the two equations for $y=1$. Interestingly, we find that for $y=1$, the two equations can be solved simulatenously only for $p \leq p_c(= 0.0454) $. This is also the value of $p$ beyond which linear stability analysis breaks down and Eq. \ref{lline} is not valid anymore. More interestingly even though the two equations can be solved for $H=0$ till $p \leq 0.0454$, we find that for $p>0.022$, one more solution shows up, with $m \neq 0$ and $H=0$. For $p>0.0454$, all possible solutions have $m \neq 0$. 

We find that the two critical lines, $\lambda_+$ and $\lambda_-$ meet $\lambda$ line at a TCP for $p<0.022$. For $p > 0.022$, the two critical lines, $\lambda_+$ and $\lambda_-$ meet inside the first order surface, i.e at a point where $m \neq 0$. This point hence is not a TCP, but a BEP. Furthermore, we find that for $p> 0.022$ there are two different kinds of phase diagrams possible: For $0.022<p \leq 0.1078$ the phase diagram is as shown in Fig. 1(b): In $H=0$ plane there is a four phase coexistence line starting from the BEP  which separates the two ordered phases. This line meets the $\lambda$-line defined via Eq. \ref{lline}, giving rise to a CEP. From CEP there is a three phase coexistence line which ends in another CEP. For $0.1078 <p \leq 0.5$ the phase diagram is as shown in Fig. 1(c): There is a four phase coexistence line from BEP which never crosses the $\lambda$ line defined via Eq \ref{lline} and goes all the way to $T=0$. Moreover, we find that $\lambda_+$ and $\lambda_-$ critical lines exist for all strengths of disorder(i.e for all values of $p$). We give more details of these multicritical points and phase topologies in the next few sections.

\section{Strong disorder and BEP}\label{sec2}
For $p>0.022$, the two critical lines for $H \neq 0$ do not meet at the potential TCP point as given by simultaneous solution of Eq \ref{lline} and 
\ref{ctcp}. Instead they meet inside the ordered plane. We find that the two wings are separated by a first order line in $H=0$ plane, which behaves differently 
for $0.022<p \leq 0.1078$ and for $0.1078 < p \leq 0.5$. Hence we will look at these two regimes separately.

\subsection{$0.1078<p \leq 0.5$}\label{sec2a}

\begin{figure}
\centering
\includegraphics[scale=0.7]{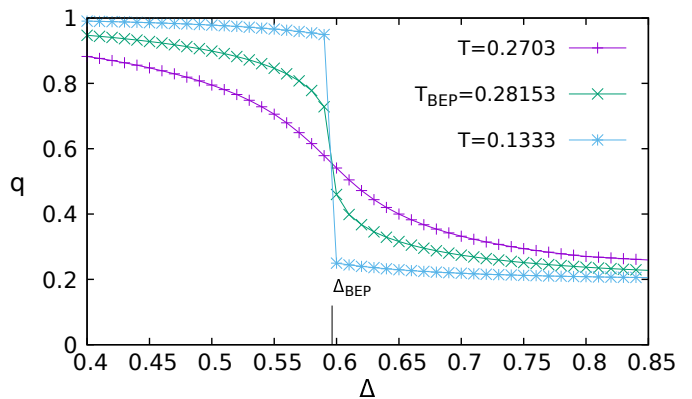}
\caption{Concentration(q) vs $\Delta$ plot for  $p=0.2$  for different values of T}
\label{conc}
\end{figure}

For this range of $p$, along the first order line in $H=0$ plane there 
is a four phase coexistence, which ends in a bicritical end point(see Fig1(c)). This line is a first order transition line between two ordered states with different values of magnetisations. These two different ordered states are a result of disorder and are not present in the pure system. At low temperatures, the system prefers $\pm 1$ spin states when $\Delta$ is small. As $\Delta$ increases, due to disorder, states with finite fraction of zero spins compete with 
the states with only $\pm 1$ spins. This can be seen by looking at the order parameter $q$ as a function of $\Delta$, as shown in Fig. \ref{conc}. 

\begin{figure}[H]
     \centering
     \begin{subfigure}[b]{0.23\textwidth}
         \centering
         \includegraphics[width=\textwidth]{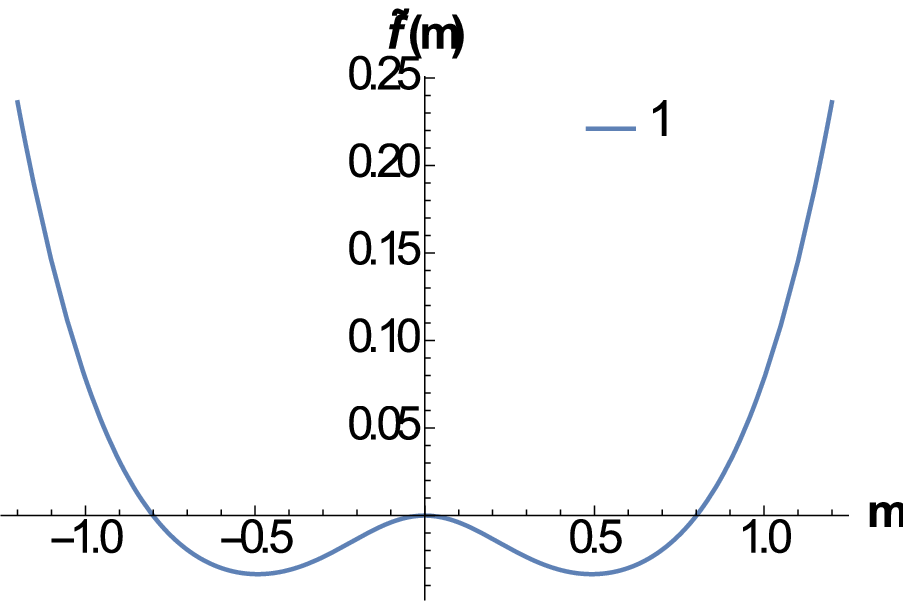}
         \label{fig:21}
     \end{subfigure}
     \hfill
     \begin{subfigure}[b]{0.23\textwidth}
         \centering
         \includegraphics[width=\textwidth]{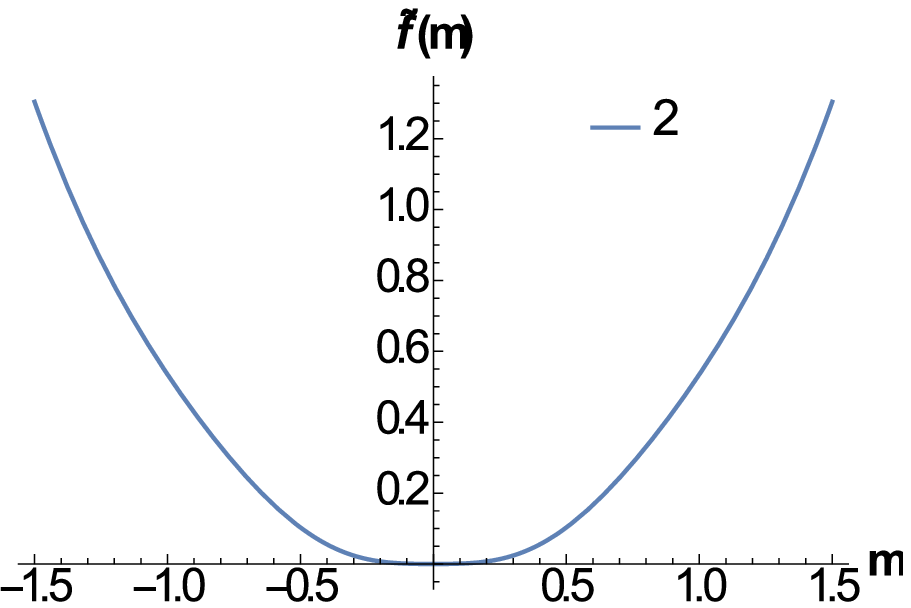}
         \label{fig:22}
     \end{subfigure}
     \hfill
     \begin{subfigure}[b]{0.23\textwidth}
         \centering
         \includegraphics[width=\textwidth]{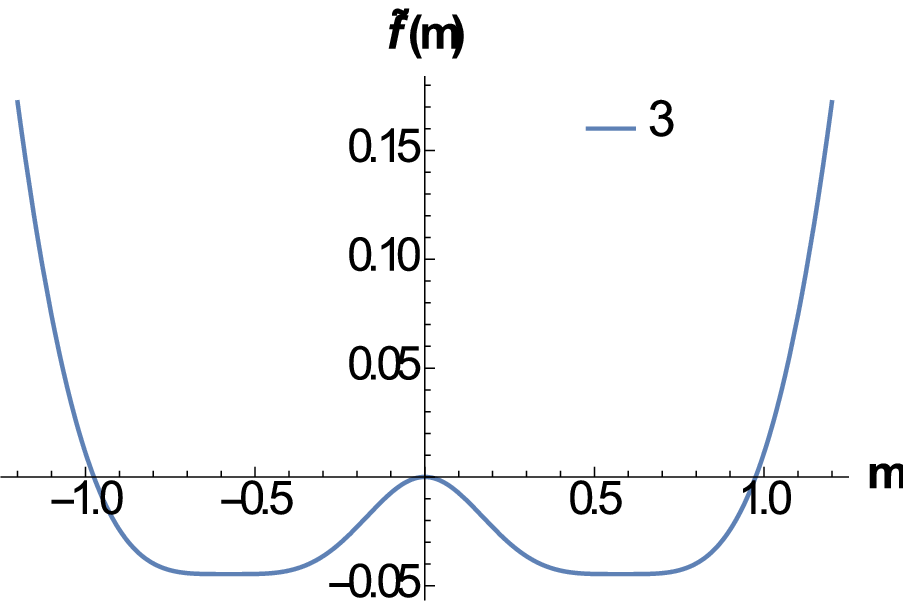}
         \label{fig:23}
     \end{subfigure}
     \begin{subfigure}[b]{0.23\textwidth}
         \centering
         \includegraphics[width=\textwidth]{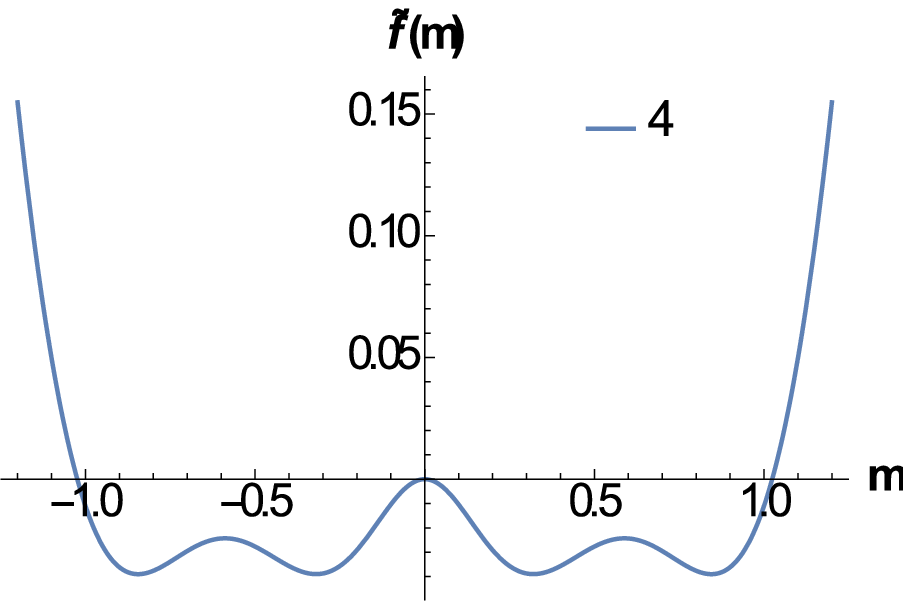}
         \label{fig:24}
     \end{subfigure}
     \hfill
     \begin{subfigure}[b]{0.23\textwidth}
         \centering
         \includegraphics[width=\textwidth]{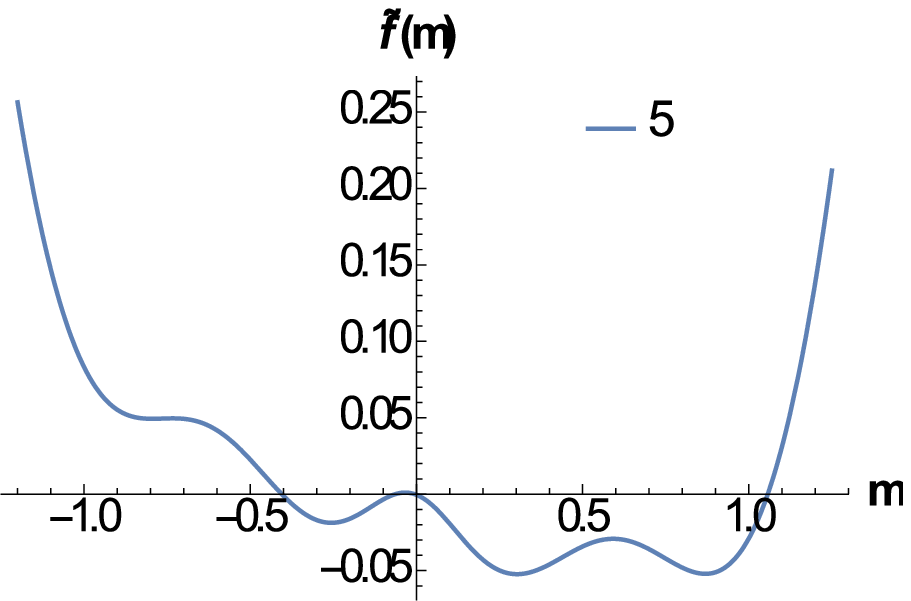}
         \label{fig:25}
     \end{subfigure}
     \hfill
     \begin{subfigure}[b]{0.23\textwidth}
         \centering
         \includegraphics[width=\textwidth]{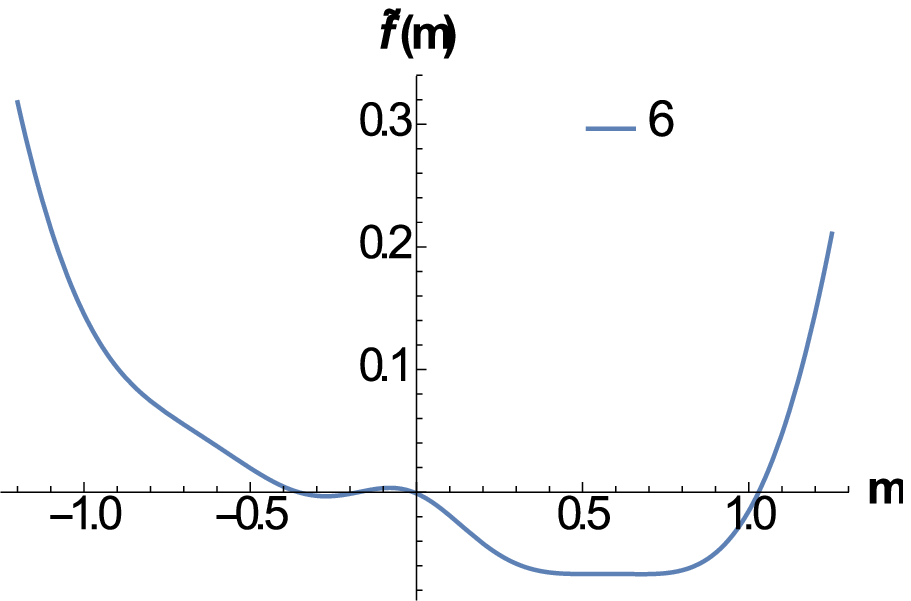}
         \label{fig:26}
     \end{subfigure}
     \begin{subfigure}[b]{0.23\textwidth}
         \centering
         \includegraphics[width=\textwidth]{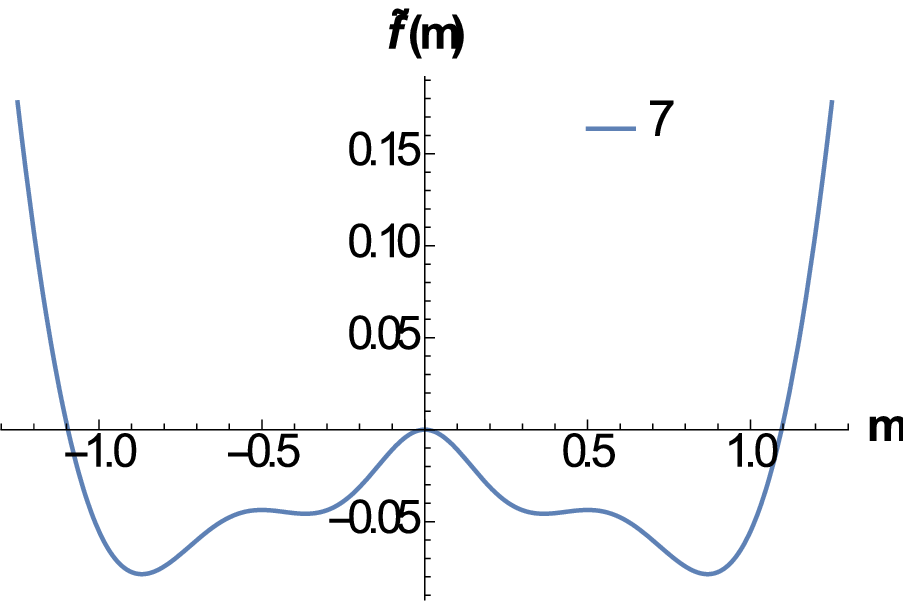}
         \label{fig:29}
     \end{subfigure}
     \begin{subfigure}[b]{0.23\textwidth}
         \centering
         \includegraphics[width=\textwidth]{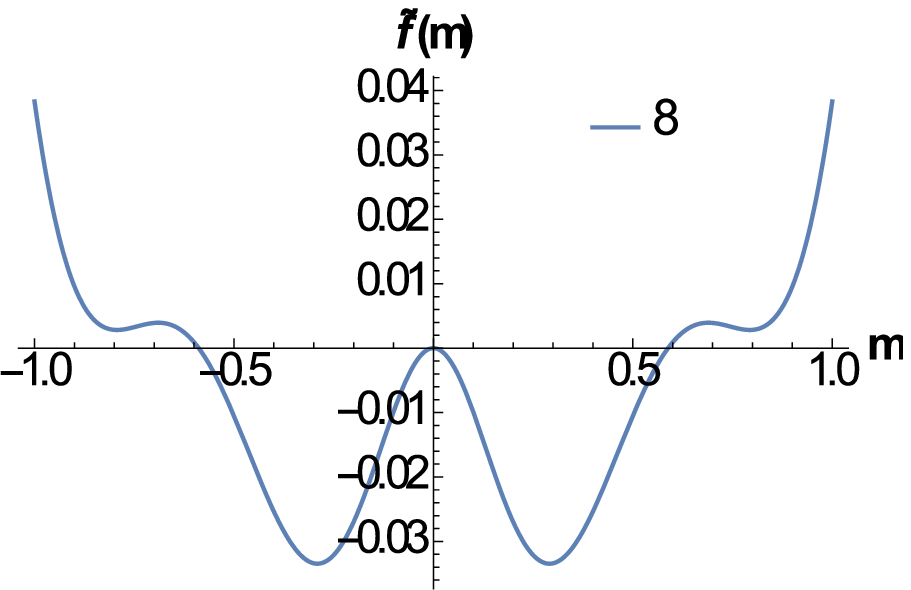}
         \label{fig:210}
     \end{subfigure}
     \hfill
             \caption{Free energy functional($\tilde{f}(m)$) as a function of $m$ in different  regions of the phase diagram(see Fig1.(c)). We have taken $p=0.2$ for which BEP is at $\Delta=0.596376$ and $T=0.2058$. The numbers on the plots refer to the numbers in Fig1(c). In (1) we plot $\tilde{f}(m)$ in $H=0$ plane just below the $\lambda$-line($T=0.27,\Delta=0.606,H=0$), in $(2)$ just above the $\lambda$ line($T=0.27,\Delta=0.864,H=0$). In (3) we show $\tilde{f}(m)$ at the BEP and one can see the coexistence of two critical phases ($T=0.2058,\Delta=0.596376,H=0$) and (4) shows the $\tilde{f}(m)$ along the quadruple coexistence line($T=0.1736,\Delta=0.59735,H=0$). In (5) we show the functional along the first order wing surface for positive $H$ ($T=0.166,\Delta=0.608,H=0.01$)and (6) shows the functional along the critical line enclosing the wing($T=0.2012,\Delta=0.615,H=0.018$). Figs (7)($T=0.1736,\Delta=0.586,H=0$) and (8)($T=0.1736,\Delta=0.61,H=0$) show the $\tilde{f}(m)$ on two sides of the first order line in $H=0$ plane.}
        \label{fig:p=2}
\end{figure}

One can see all the transitions clearly by plotting $\tilde{f}(m)$ in different regions of the phase diagram as shown in Fig. \ref{fig:p=2} for $p=0.2$. From the plots we can see that the $H=0$ line separates the two ordered phases. Along $H \neq 0$ critical lines, two of these phases become critical and at BEP the two critical phases coexist. 

\begin{figure}
\centering
\includegraphics[scale=0.7]{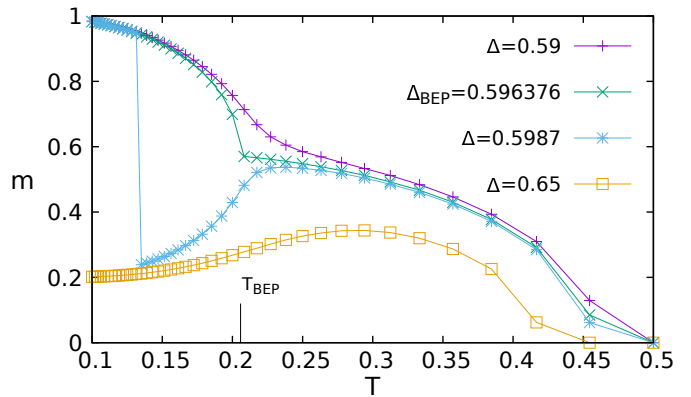}
\caption{Magnetization(m) vs $T$ plot for  $p=0.2$ for different values of $\Delta$ for $H=0$. At BEP the first order jump vanishes and near $T=T_{BEP}$ one sees a change in slope for broad range of $\Delta$. }
\label{mbep1}
\end{figure}

\begin{figure}
\centering
\includegraphics[scale=0.7]{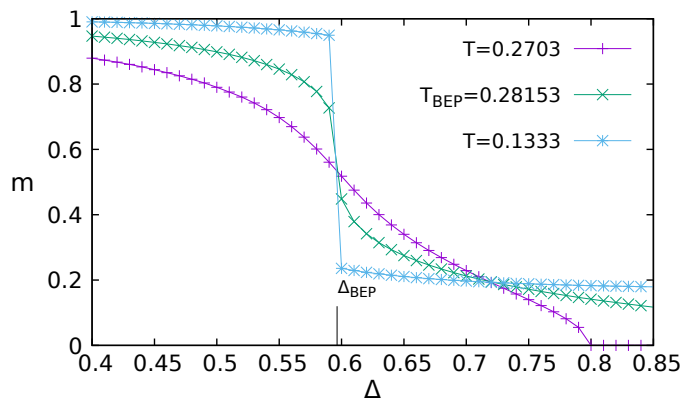}
\caption{Magnetization(m) vs $\Delta$ plot for $p=0.2$ for different values of $T$.}
\label{mbep2}
\end{figure}

To understand the nature of transition especially at 
BEP, we looked at the magnetisation$(m)$ and magnetic susceptibility $(\chi)=\frac{\partial m}{\partial H} |_{H \rightarrow 0}$. Let us first look at the magnetisation as a function of $T$ in the $H=0$ plane for different fixed values of $\Delta$ (see Fig. \ref{mbep1}). We find that for $\Delta<\Delta_{BEP}$, the magnetisation changes its slope near $T=T_{BEP}$, the change becomes sharper as one approaches $\Delta=\Delta_{BEP}$. For $\Delta > \Delta_{BEP}$(but close to $\Delta_{BEP}$),  the magnetisation undergoes a first order transition as it crosses the quadruple line and then changes slope near $T=T_{BEP}$. For $\Delta$ much larger than $\Delta_{BEP}$, as we increase $T$ there is no first order jump or change of slope around $T=T_{BEP}$. We also looked at $m$ as a function of $\Delta$ for three different values of $T$ (see fig. \ref{mbep2}). First order jump as one crosses the quadruple line is clear for $T<T_{BEP}$. For $T>T_{BEP}$ there is no signature of any transition.

\begin{figure}
\centering
\includegraphics[scale=0.7]{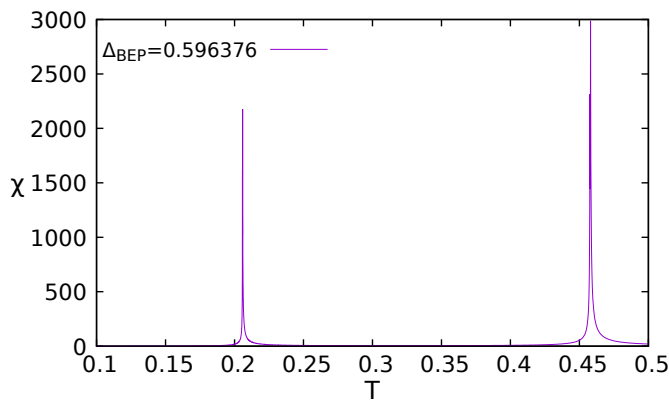}
\caption{Magnetic susceptibility($\chi$) vs $T$ plot at $\Delta_{BEP}$ 
for  $p=0.2$ }
\label{msbepd}
\end{figure}

\begin{figure}
\centering
\includegraphics[scale=0.7]{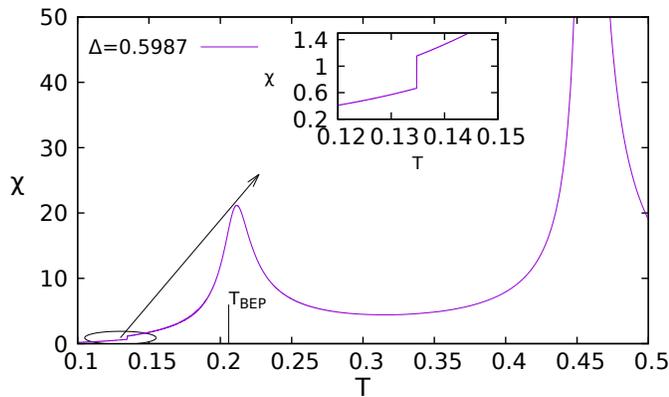}
\caption{Magnetic susceptibility($\chi$) vs $T$ plot for  $p=0.2$ for $\Delta >\Delta_{BEP}$ }
\label{msgbepd}
\end{figure}

It is hard to deduce the nature of transition at BEP by looking at the  magnetisation alone. Hence we studied the magnetic susceptibility near BEP. First we look at it for fixed value of $\Delta$. As we fix $\Delta=\Delta_{BEP}$ and vary $T$, we find that there is an infinite peak at the $T$ of $\lambda$ transition. There is another peak at $T=T_{BEP}$, but this peak is finite(see Fig. \ref{msbepd}).   This behaviour can be contrasted with the behaviour at $\Delta> \Delta
_{BEP}$ as shown in Fig \ref{msgbepd}. We find a discontinuity where it crosses the first order line and a finite peak near $T=T_{BEP}$.

We also studied magnetic susceptibility as we vary $\Delta$ at $T_{BEP}$. As expected, we found that there is a finite peak at $\Delta=\Delta_{BEP}$ For $T<T_{BEP}$, there was instead a first order jump in magnetic susceptibility. 
We scanned a large region in $(T,\Delta)$ plane near BEP. We find that the effect of the presence of BEP is felt even far away from the point. But the magnetisation and susceptibility plots are smooth near BEP(though susceptibility shows a cusp). It was shown via scaling arguments \cite{helena-barbosa} that if the two critical lines meeting at BEP are in the same universality class and are symmetric, then the singular behaviour contribution to the phase boundary cancels out \cite{helena-barbosa, plascak}. In our case 
the two critical lines $\lambda_+$ and $\lambda_-$ lie in the Ising  universality class. Looking at the three dimensional phase diagram it is clear that there is only one phase in the system in the sense that there exist a path between any two non-singular points in the phase diagram which does not  have to encounter a singularity. At BEP the first three derivatives of $\tilde{f}(m)$ w.r.t $m$ are zero and hence the free energy is not analytic at this point. Hence, we conclude that BEP is a point of two phase co-existence and there is no critical transition from one phase to another at BEP.

 As $p$ increases we find that the critical lines enclosing the wings become flatter and the temperature at which they meet in $H=0$ plane decreases. We have tabulated the range of $T$ for different $p$ in Table 1. 

\begin{table*}[ht]
\begin{center}
\begin{tabular}{|c|c|c|c|c|}
\hline
\multicolumn{5}{|c|}{$0.022 <p \leq 0.5$}\\
\hline
$p$ & $T_{lc}$ & $\Delta_{lc}$ & $T_{uc}$  & $\delta T$  \\
\hline
0.0453 &0.28043 &0.501175 &0.23866    &0.0417665   \\
\hline
0.05 &0.276396 &0.50468 &0.237473  &0.038923  \\
\hline
0.07 &0.26185 &0.518896 &0.23245  &0.029399 \\
\hline
0.1 &0.2451 &0.538417 &0.224972   &0.020128 \\
\hline
0.2 & 0.2058 & 0.596376  &0.2  &0.0058 \\
\hline
0.3 &0.17643 &0.6490843 &0.174978  &0.001452  \\
\hline
0.4 &0.15024 &0.69968 &0.1499  &0.000248  \\
\hline
0.5 &0.125016 &0.7499884 &0.12498	  &0.000036  \\
\hline

\end{tabular}
\end{center}
\caption{Width of the wing lines for different $p$. $T_{lc}$ and $\Delta_{lc}$ represent the values of $T$ and $\Delta$ for $H=0$ where the $\lambda_+$ and $\lambda_-$ lines meet and $T_{uc}$ is the value along the critical line as $\Delta \rightarrow \infty$ and $H \rightarrow \infty$. }
\label{tab:1}
\end{table*}
 
\subsection{$0.022<p \leq 0.1078$}\label{sec2b}
\begin{figure}
     \centering
     \begin{subfigure}[b]{0.23\textwidth}
         \centering
         \includegraphics[width=\textwidth]{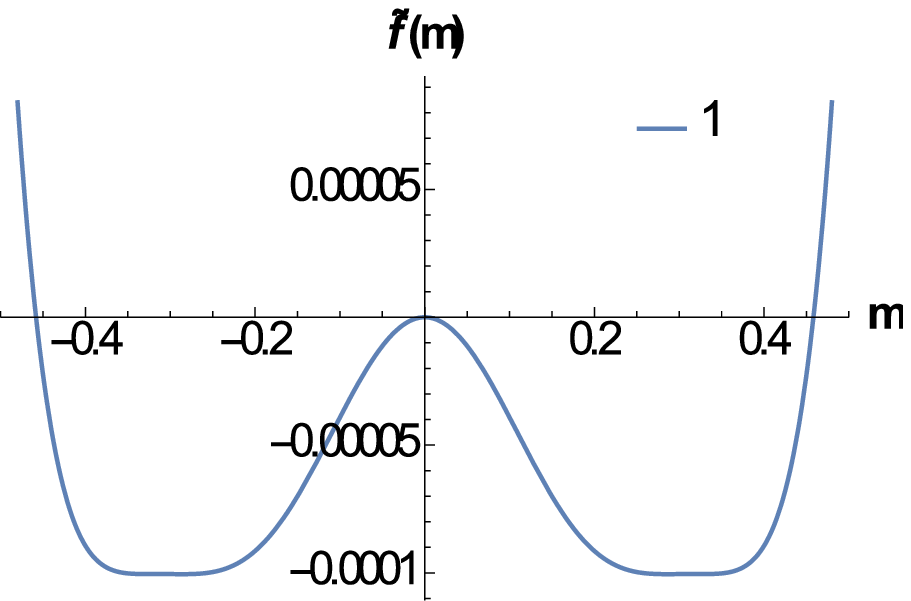}
         \label{fig:443}
     \end{subfigure}
     \begin{subfigure}[b]{0.23\textwidth}
         \centering
         \includegraphics[width=\textwidth]{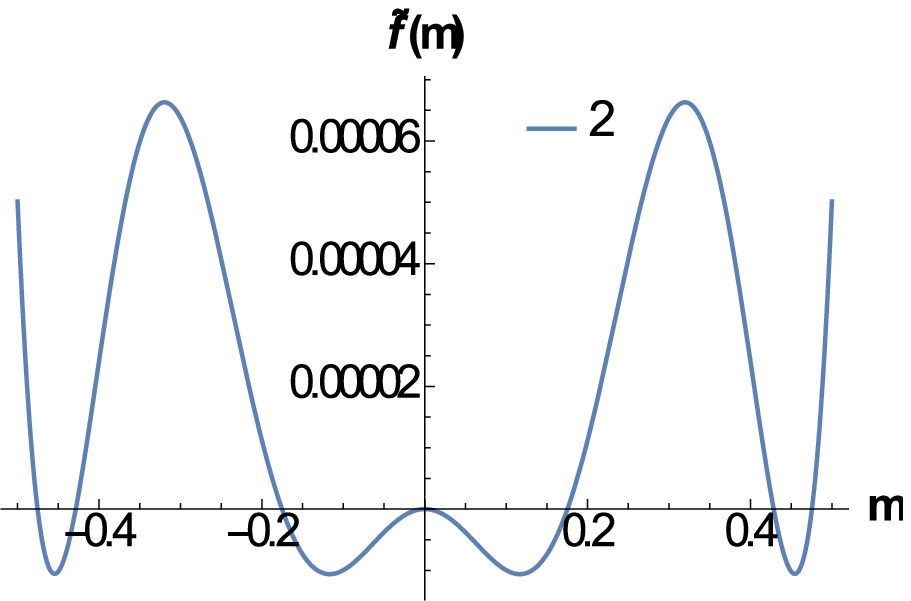}
         \label{fig:444}
     \end{subfigure}
     \hfill
     \begin{subfigure}[b]{0.23\textwidth}
         \centering
         \includegraphics[width=\textwidth]{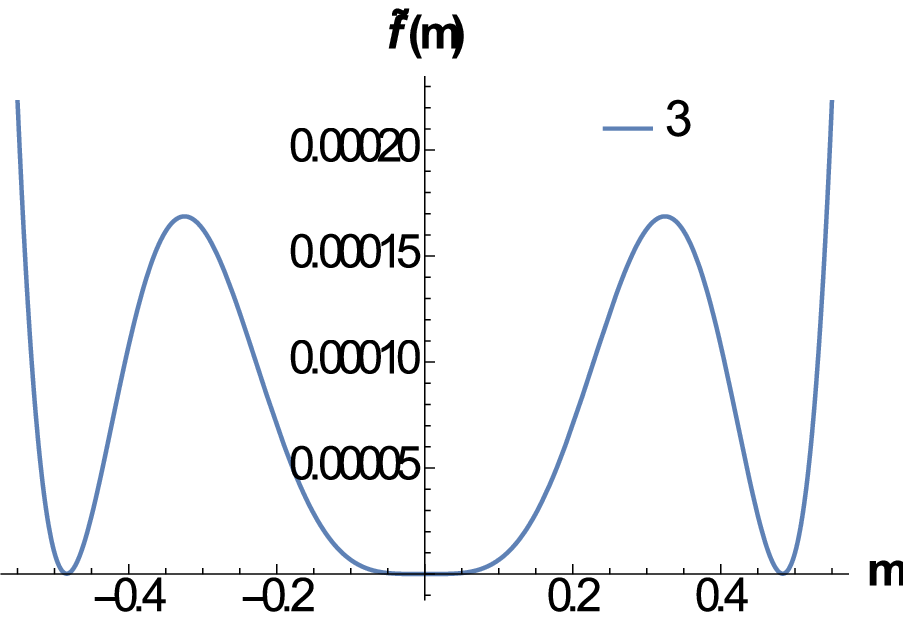}
         \label{fig:449}
     \end{subfigure}
     \hfill
     \begin{subfigure}[b]{0.23\textwidth}
         \centering
         \includegraphics[width=\textwidth]{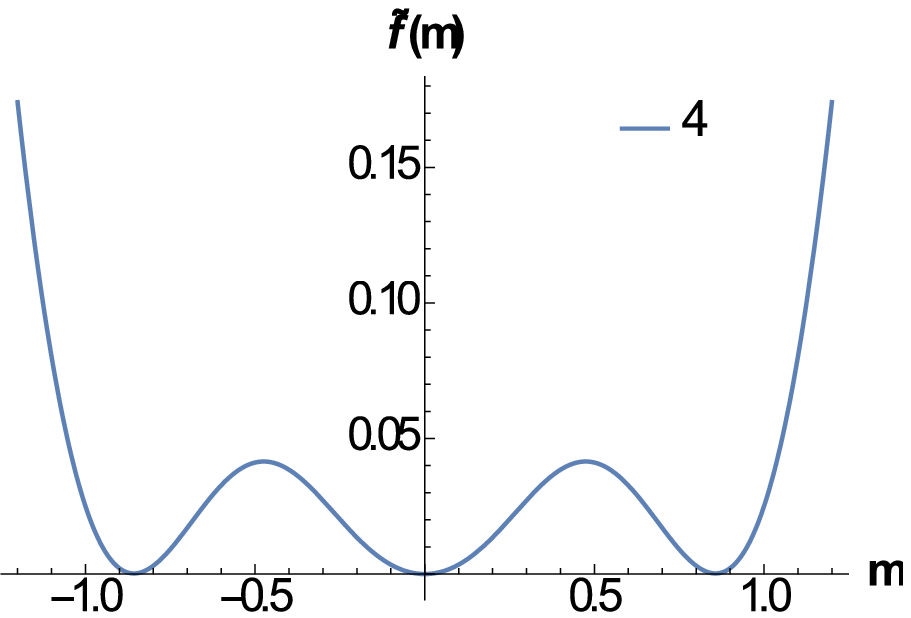}
         \label{fig:4410}
     \end{subfigure}
     \hfill
     \begin{subfigure}[b]{0.23\textwidth}
         \centering
         \includegraphics[width=\textwidth]{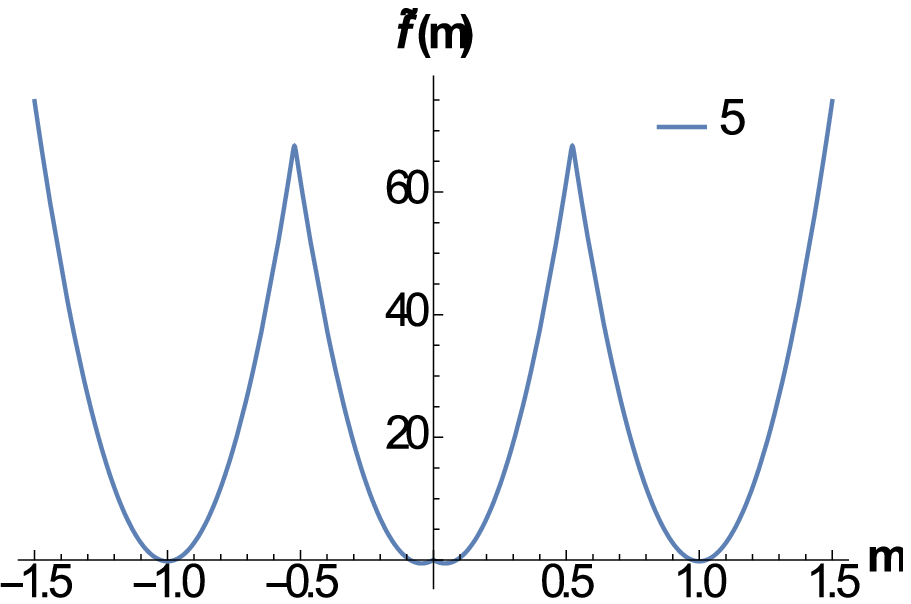}
         \label{f}
     \end{subfigure}
     \hfill  
        \caption{Free energy functional $\tilde{f}(m)$ plots for  $p=0.0044$. (1) $\tilde{f}(m)$ at the BEP with $T=0.281532,\Delta=0.500195, H=0$;(2)$\tilde{f}(m)$  along the first order line between BEP and CEP1($T=0.2777,\Delta=0.500183,H=0$). (3) $\tilde{f}(m)$ at the CEP1 with $T=0.27585, \Delta=0.500186$,$H=0$, (4) shows the functional along the first order line from CEP1 to CEP2($T=0.2,\Delta=0.5088,H=0$), (5) shows functional along the first order line from CEP2 at $T=0.02$ to $T=0$(at $T=0.001667,\Delta=0.5226,H=0$)}
        \label{fcep}
\end{figure}

In this region the wings meet at BEP as before, but the first order quadruple line now intersects the $\lambda$-line at a critical end point(we will call this critical end point as CEP1 to distinguish it from the other critical end point in the phase diagram at a lower temperature, which we will call as CEP2). After that it becomes a line of triple point(see Fig1(b)). In Fig.\ref{fcep},  we plot the free energy functional along this line. Along the first order line there is a line of four phase coexistence between BEP and CEP1 and then there is a usual triple line between CEP1 and CEP2. As shown in Fig. \ref{fcep}(3), CEP1 itself is neither a quaduple or a triple point. It is instead a point where a critical state coexists with two ordinary stable phases. Between CEP2 and $0$ temperature there is again a quadruple line as shown in Fig \ref{fcep}(5).

The CEP is a point where two phases become critical in the presence of one or more non critical spectator phase. At CEP, $\tilde{f}(m)$ for $m=0$ and for $m \neq 0$ should be equal (i.e $\tilde{f}(m=0)=\tilde{f}(m \neq 0)$) along with their derivative with respect to $m$ ($\tilde{f}'(m=0)=\tilde{f}'(m \neq 0)$). If this point lies on the $\lambda$ line, then we get the condition for CEP. Hence to find CEP, we explore the $\lambda$-line for a point where $\tilde{f}(m=0)=\tilde{f}(m \neq 0)$  along with $\tilde{f}'(m=0)=\tilde{f}'(m \neq 0)$. We find that for $p> 0.1078$ the condition cannot be satisfied. 

\begin{figure}
     \centering
     \begin{subfigure}[b]{0.23\textwidth}
         \centering
         \includegraphics[width=\textwidth]{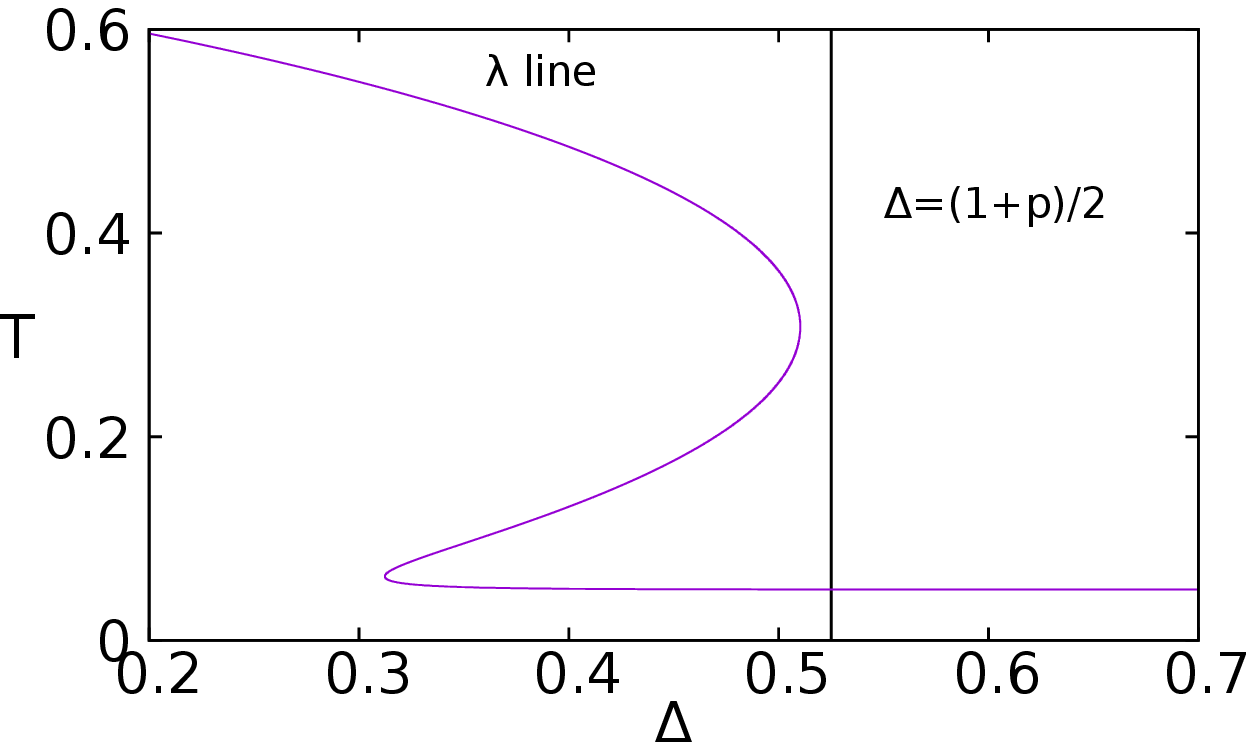}
         \caption{}
         \label{fig:443}
     \end{subfigure}
     \begin{subfigure}[b]{0.23\textwidth}
         \centering
         \includegraphics[width=\textwidth]{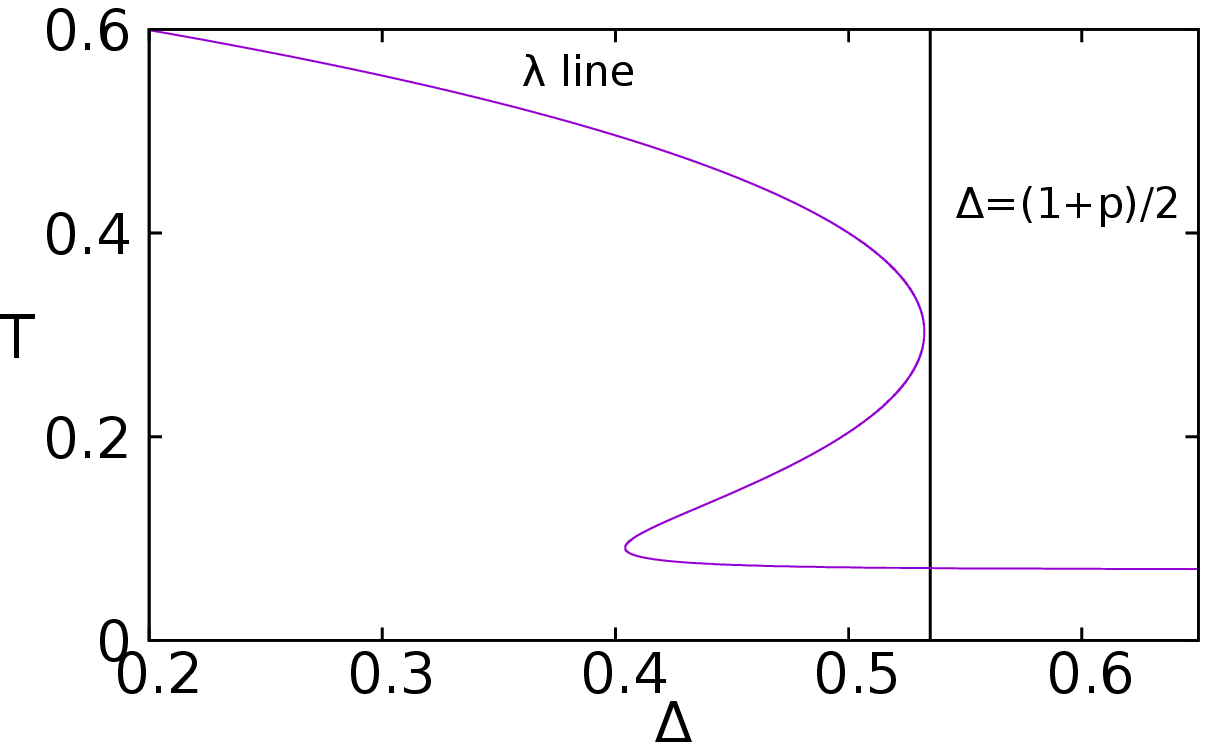}
         \caption{}
         \label{fig:444}
     \end{subfigure}
     \hfill
     \begin{subfigure}[b]{0.23\textwidth}
         \centering
         \includegraphics[width=\textwidth]{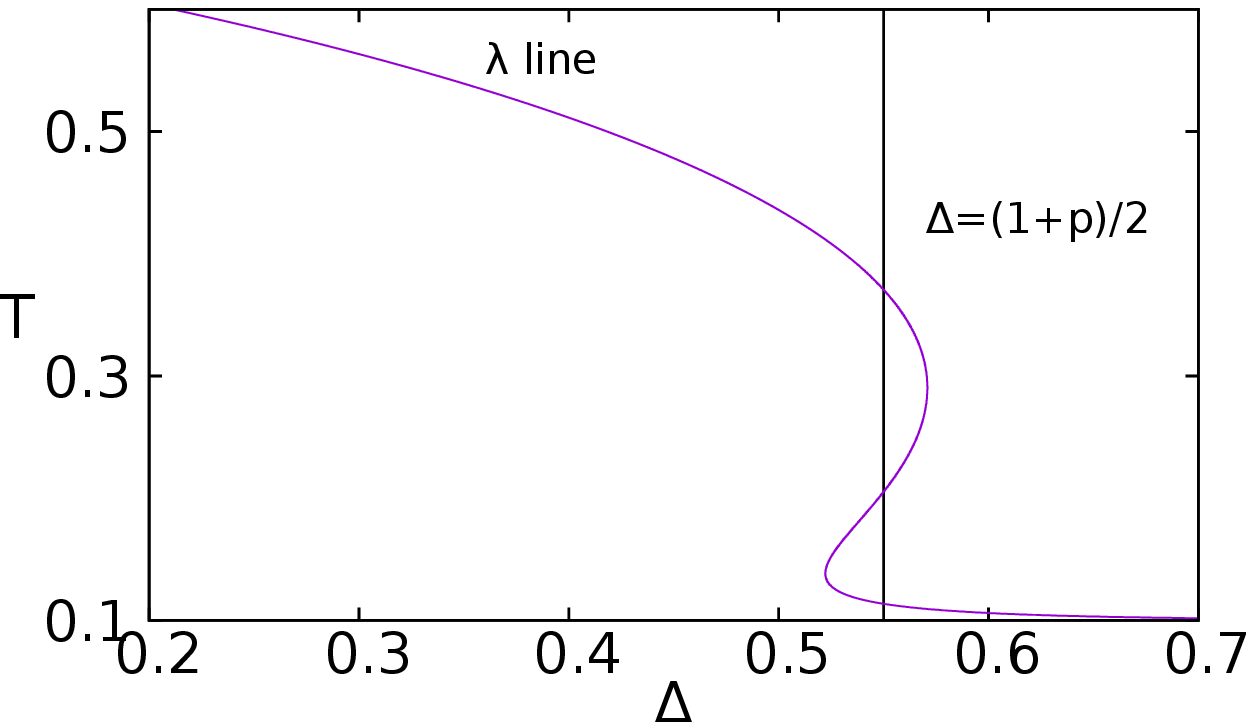}
         \caption{}
         \label{fig:449}
     \end{subfigure}
     \hfill
     \begin{subfigure}[b]{0.23\textwidth}
         \centering
         \includegraphics[width=\textwidth]{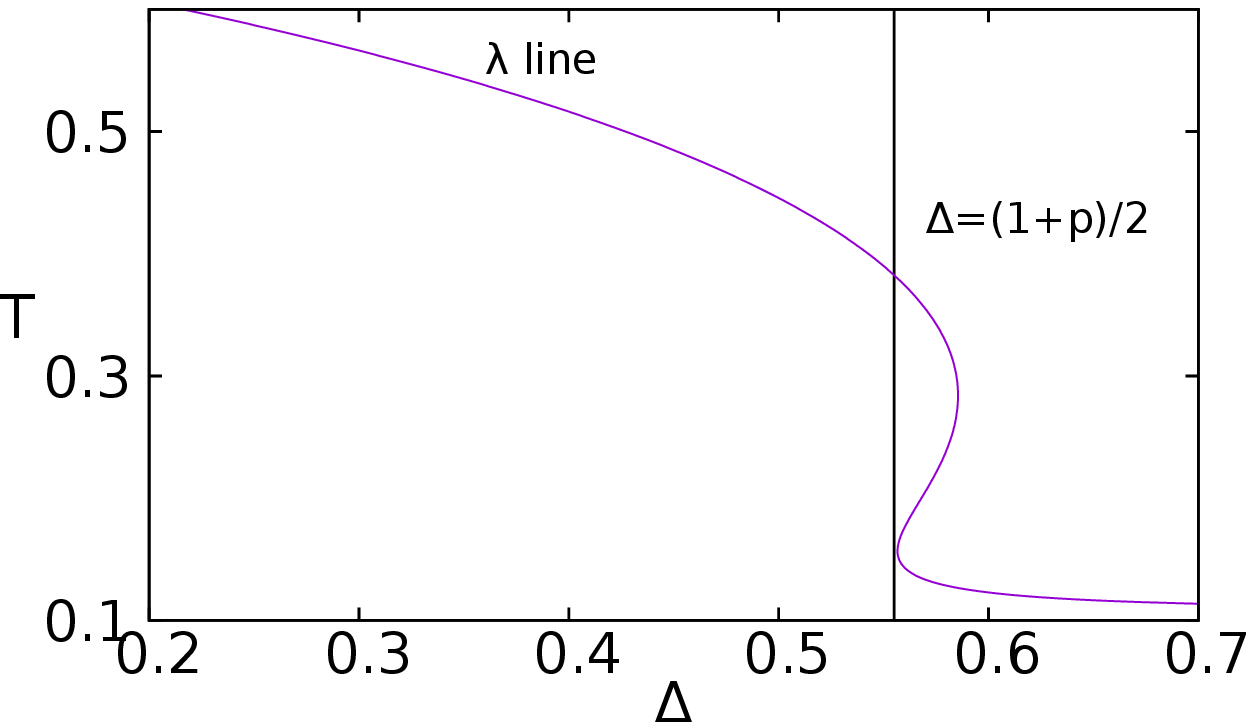}
         \caption{}
         \label{fig:4410}
     \end{subfigure}
     \hfill
        \caption{$\lambda$-line plotted along with $\Delta=(1+p)/2$. (a) $p=0.05$,(b)$p=0.07$,(c)$p=0.10$ and (d)$p=0.11$. For $p<=0.07$ the $\Delta=(1+p)/2$ line intersect the $\lambda$-curve only once. For $0.07<p<0.11$ it intersects it three times and only for $p>=0.11$ it is fully on the left of the curve and hence doesn't intersect}
        \label{pcep}
\end{figure}

In \cite{santos}, Santos et al also reported the presence of CEP1 and CEP2 for $0.022 < p < 0.074$, by looking at the point of intersection of the $\lambda$-line with the first order line. We find that this topology extends till $p=0.1078$. In order to understand the discrepancy, we have plotted $\lambda$-line given by Eq. \ref{lline} along with a line parallel to $T$-axis at $\Delta=(1+p)/2$ in Fig. \ref{pcep}. The line $\Delta =(1+p)/2$ is a good approximation to the first order line in the $(T,\Delta)$ plane as we found that the first order line is almost parallel to $T$-axis. As shown in Fig. \ref{pcep}, $\Delta=(1+p)/2$ line crosses the $\lambda$-line once till $p \approx 0.07$ and thrice for $0.07<p<0.11$. For $p>=0.11$ there is no intersection. More careful analysis using the full free energy functional, gives us the value to be around $p=0.1078$. This matches with the value obtained by equating the free energy functional and its first derivative along the $\lambda$-line, as described in the previous paragraph.
 
In Table 2 we tabulate the location of BEP, CEP1 and CEP2 for different values of $p$. The first order line between CEP2 and $\Delta$- axis is similar to the first order line reported in Sec \ref{sec2a}, which separates the states with almost all $\pm 1$ spins from a state with 
$p$ fraction of $\pm 1$ spins. The presence of CEP1 and a four phase co-existence line between BEP and CEP1 is due to the occurence of a new magnetic state. This state has more than $p$ fraction of $\pm 1$ spins, as it occurs at a higher temperature, very close to the $\lambda$ line. 

\begin{table*}[ht]
\begin{center}
\begin{tabular}{|c|c|c|c|c|c|c|}
\hline
\multicolumn{7}{|c|}{$0.022 < p \leq 0.107578$}\\
\hline
$p$ & $T_{BEP}$ & $\bigtriangleup_{BEP}$ & $T_{CEP1}$ & $\bigtriangleup_{CEP1}$& $T_{CEP2}$ &  $\bigtriangleup_{CEP2}$  \\
\hline

0.03 &0.2961208 &0.489187 &0.295197 &0.489166 &0.03 &0.4977229  \\
\hline
0.044 &0.28153 &0.500195 &0.27585 &0.500186 &0.04401 &0.521585  \\
\hline
0.07 &0.26185 &0.518896 &0.24036 &0.519398 &0.07099 &0.533953 \\
\hline
0.107 &0.24166 &0.542 &0.15972 &0.547514 &0.13975 &0.549068  \\
\hline

\end{tabular}

\end{center}
\caption{Co-ordinates of the BEP and CEP's for $0.022<p<0.107$.}
\label{tab:2}
\end{table*}

\begin{figure}
\centering
\includegraphics[scale=0.7]{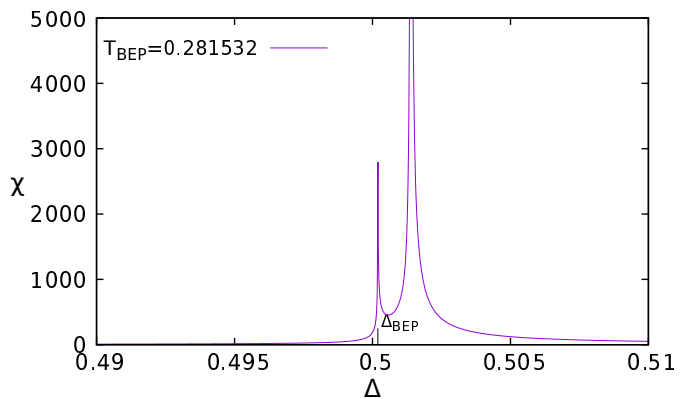}
\caption{Magnetic susceptibility($\chi$) vs $\Delta$ plot for  $p=0.044$ at $T=T_{BEP}$}
\label{wbep}
\end{figure}

\begin{figure}
\centering
\includegraphics[scale=0.7]{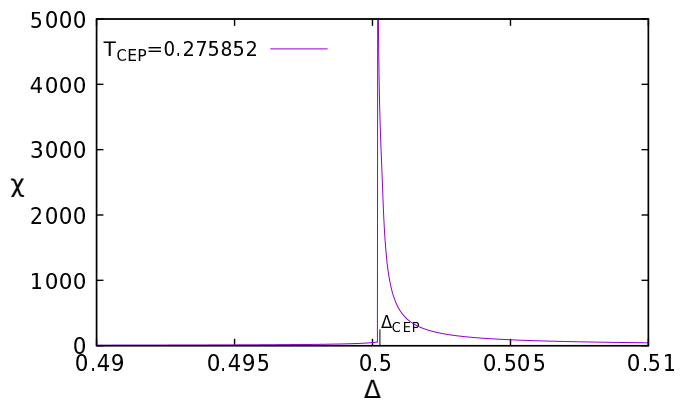}
\caption{Magnetic susceptibility($\chi$) vs $\Delta$ plot for  $p=0.044$ at $T=T_{CEP}$}
\label{wcep}
\end{figure}

We plot the magnetic susceptibility as a function of $\Delta$ for 
$T=T_{BEP}$ and for $T=T_{CEP}$ in $H=0$ plane. As expected first 
plot shows two peaks:a finite peak at $BEP$ and an infinite peak at intersection with the $\lambda$-line (see Fig. \ref{wbep}), while the second plot shows one peak only at CEP1 (see Fig. \ref{wcep}).

\section{Weak disorder and TCP}\label{sec3}
Along the region $0 \leq p \leq 0.022$ the wings meet the $\lambda$-line at the TCP and the phase diagram is similar to the pure case. Along the first order line there is three phase coexistence. As $p$ increases, the TCP shifts towards smaller $T$ and larger $\Delta$. At $p=0.022$ the TCP becomes a fourth order critical point. Again, at very low temperature there is a CEP, similar to the case discussed in the Section \ref{sec2b} for all $p>0$, which is a state with $p$ fraction of $\pm 1$ spins and occurs at low temperatures on a complete graph. For $0.017<p<0.022$, there is re-entrance region in the phase diagram, as the TCP does not coincide with the maximum of $\lambda$-line given by Eq. \ref{lline}.
\section{Landau theory}\label{sec4}
In the previous sections we studied the phase diagram by looking at the full free energy functional and its derivatives. Usually, Landau theory is a very useful tool to classify different kinds of transitions and even though it might not be accurate quantitatively, it helps in understanding different possible topologies of the phase diagram. But while very successful in explaining ordinary critical point, it is not always possible to find a Landau description for higher order critical point, i.e. it is perhaps possible to define a free functional always, but it might not always be Taylor expandable \cite{deepak}. In this section we expand the free energy functional to check if we can explain the phase diagrams based on the  coefficients of different powers of the order parameter. For example, the Ising universality class critical point can be determined easily by expanding upto fourth power in $m$, provided that the next higher order coefficient is positive. For TCP one needs to expand till sixth order. A sixth order Landau theory hence allows only for ordinary critical points and TCPs. We expect that we need to keep more terms in the expansion, if we expect to find higher order critical points like CEP and BEP \cite{japan}. Hence we expanded the free energy functional till eighth power of $m$. We get
\begin{equation} \label{eq:16}
\tilde{f}(m)=a_2 m^2 + a_4 m^4+ a_6 m^6+ a_8 m^8
\end{equation}
where $a_i$'s are Landau coefficients, as follows:
\begin{eqnarray*}\label{coefficients}
&& a_2 =\frac{\beta}{2} \Bigg (1 + \frac{2 \beta  (p-1)}{2 + e^{\beta \Delta}} - \frac{2 \beta  p e^{\beta \Delta}}{1 + 2 e^{\beta \Delta}} \Bigg) \nonumber \\
&& a_4 = \frac{\beta^4}{12}  \Bigg(\frac{(-4 + e^{\beta \Delta})(p-1)}{(2 + e^{\beta \Delta})^2} + \frac{p e^{\beta \Delta}(-1 + 4 e^{\beta \Delta})}{(1 + 2 e^{\beta \Delta})^2 }
    \Bigg) \nonumber \\
&& a_6 = \frac{\beta^6}{360}  \Bigg(\frac{(64 - 26 e^{\beta \Delta} + e^{2 \beta \Delta})(p-1)}{(2 + 
     e^{\beta \Delta})^3}  \nonumber \\
     && \; \; \; \; \;- \frac{p e^{\beta \Delta} (1 - 26 e^{\beta \Delta} + 64 e^{2 \beta \Delta}) }{
  (1 + 2 e^{\beta \Delta})^3} \Bigg) \nonumber \\
\end{eqnarray*}
\begin{eqnarray}
 && a_8 = \frac{\beta^8}{20160} \Bigg (\frac{( 1188 e^{\beta \Delta}-2176  - 120 e^{2 \beta \Delta} + e^{
      3 \beta \Delta}) (p-1)}{(2 + e^{\beta \Delta})^4} \nonumber \\
      && \; \; \; \; \;+ \frac{p 
  e^{\beta \Delta} (-1 + 120 e^{\beta \Delta} - 1188 e^{2 \beta \Delta} + 
      2176 e^{3 \beta \Delta})}{(1 + 2 e^{\beta \Delta})^4} \Bigg)
\end{eqnarray}

The second order transition is given by $a_2=0$, provided $a_4 >0$. Equating $a_2=0$ gives us:
\begin{equation}\label{llline}
1 + \frac{2 \beta  (p-1)}{2 + e^{\beta \Delta}} = \frac{2 \beta p e^{\beta \Delta}}{1 + 2 e^{\beta \Delta}} 
\end{equation}
This equation is same as Eq. \ref{lline}, obtained by linear expansion around $m=0$. According to the Landau theory, a new universality class, namely the TCP occurs when $a_4$ becomes equal to $0$, provided $a_6>0$. We find that the condition for 
$a_4=0$ along the $\lambda$-line is the same as given by  substituting  Eq. \ref{ctcp} into Eq. \ref{lline}. For $p>p_c=0.0454$, $a_4$ is never $0$ and hence beyond $p_c$ the condition for occurence of TCP cannot be satisfied. For $p>0.022$, $a_6<0$ at the point where $a_4=0$. Hence sixth order Landau theory while sufficient for $p<0.022$, is not enough for $p>0.022$. 

Hence for $a_6<0$, we consider the expansion till eighth order, since $a_8>0$ for all ranges of the parameters. CEP will be a point along the $\lambda$-line (given by Eq. \ref{llline}) where the $\tilde{f}(T_c,m_c)=0$ and $\tilde{f}'(T_c,m_c)=0$ and $m_c \neq 0$. Solving these, we get the condition for the existence of CEP to be
\begin{equation}\label{lcep}
\frac{a_6^2}{4 a_4 a_8}=1
\end{equation}
We find that Eq. \ref{lcep} can be satisfied only for $0.022<p \leq 0.0454$, and that too at a point very close to the point where $a_4=0$. For example, for $p=0.044$ from Eq. \ref{lcep}, we get $(T_{CEP1},\Delta_{CEP1}) =(0.267,0.497)$ and for $p=0.03$ we get $(T_{CEP1},\Delta_{CEP1}) =(0.294,0.489)$. Hence we find that the value obtained 
via Eq. \ref{lcep} are different from the ones obtained by looking at the full free energy functional in Section \ref{sec2b}(see Table II). The difference increases with increasing $p$. More importantly, in Section \ref{sec2b} we had found numerically that CEP is present for a much larger range of $p$: $0.022<p \leq 0.1078$. 

To estimate BEP using truncated $\tilde{f}(m)$, we equate the first three derivatives of the truncated $\tilde{f}(m)$ in Eq. \ref{eq:16} w.r.t $m$ to $0$. For $m \neq 0$, this gives the condition for BEP to be: 
$a_6=-\sqrt{\frac{8 a_4 a_8}{3}}$. Again this condition gets satisfied only for $0.022<p \leq 0.0454$. This gives a BEP very close to CEP and the actual location does not match with the numerical estimates of Section \ref{sec2}. Hence, a Landau description of this system predicts the phase diagram correctly for $p<0.022$(except for CEP present at very low temperatures for all $p>0$) and gives qualitatively similar diagram for $0.022<p<0.0454$, though the location of BEP and CEPs does not match the actual value. For $p >0.0454$ it is inadequate in predicting the phase diagram. We tried including more terms in the expansion of $\tilde{f}(m)$, but we could not locate BEP using a truncated $\tilde{f}(m)$, suggesting that full $\tilde{f}(m)$ is needed for locating the BEP.

\section{Discussion}\label{sec5}

Blume-Capel model is a very useful model due to its simplicity and 
rich phase diagram. Its phase diagram in the presence of 
disorder in $(T,\Delta)$ plane has been studied extensively using 
many different techniques. In this paper we studied the three 
field phase diagram in the presence of disorder, which has not 
been studied earlier. This is useful especially to correctly 
predict the nature of multicritical points. We found as the 
disorder strength increases, the two wings meet at a BEP. We 
showed that this is actually a point of co-existence of two 
critical phases, where the magnetic susceptibility is finite. 
Hence in-spite of the three derivatives of the free energy being 
zero at BEP, it is not critical. This point was identified as an 
ordered critical point in earlier studies \cite{santos, salmon}. 
Also, we corrected the estimate of onset of topology III as a 
function of disorder strength compared to earlier estimate \cite{santos}. In \cite{mfdilution} a different bimodal
 distribution of the random crystal field ($P(\Delta_i)= p \delta(\Delta_i-\Delta)+(1-p) \delta(\Delta)$) was studied using 
Landau theory. They observed that as the disorder anisotropy 
increases, the first order line meets the second order line 
such that there is a re-entrant part in the phase diagram. By analogy with binary fluids, they conjectured that when the second order line has a re-entrant part, it will end in a double critical point(or bicritical end point) followed by a CEP as in Fig 1(b). Since their work was based on Landau expansion, they could not identify the bicritical end point and critical end point precisely. Also they did not report a topology similar to Fig. 1(c) for strong disorder.

It would be interesting to see if similar phase diagrams are realized in finite dimensions using numerical simulations \cite{fytas,deserno,sethna}. The origin of BEP in our model is different than that for the pure anisotropic continuous spin systems, where the BEP was seen as an end-point of spin flop transition line \cite{nelson}. A two parameter Landau theory description exists for spin-flop \cite{chaikin}. Our free energy functional is a one parameter function which shows BEP. It would be interesting to see if a one parameter Landau theory can be built, which has a BEP as seen in the strong disorder case in our work. A study of three fields diagram for random field Blume Capel model would also be useful to understand the nature of TCPs reported in \cite{rf} in the $(T,\Delta)$ phase diagram \cite{wip}.

\section{Acknowledgement}

We thank Deepak Dhar for valuable comments on the manuscript and Mustansir Barma for discussions. S thanks ICTP for hospitality during the completion of this work.


\begin{thebibliography}{99}
\bibitem{knobler} C. M Knobler and R. L. Scott, "Multicritical points in fluid mixtures",Editors:C. Domb and J. L. Lebowitz, Vol. 9, Academic Press, New York 1984.
\bibitem{fisher} Michael E.Fisher, Critical endpoints, interfaces, and walls, Physica A{\bf 172} 77-86 (1991).
\bibitem{griffithsh} R. B. Griffiths, Phys. Rev. B, {\bf 12} 345 (1975).
\bibitem{widom}  B Widom, Journal of Chem. Phys., {\bf 67}, 872(1977).
\bibitem{binarypd} I Nezbeda, J Kolafa and W. R Smith, J. Chem. Soc. Faraday Trans., {\bf 93},3073,1997.
\bibitem{metamagnet} E Stryjewski and N. Giordano, Adv. in Phys. {\bf 26}487(1977).
\bibitem{alloys} D. P. Lara, G. A. P. Alcázar, L. E. Zamora, and J. A. Plascak, Phys. Rev. B, {\bf 80} 014427(2009).
\bibitem{helium} C. Buzano, M. Cieplak, M. R. Swift, F. Toigo and J. R. Banavar, Phys. Rev. Lett. {\bf 69} 221(1992)
\bibitem{kirkpatrick} D Belitz, T. R. Kirkpatrick and J Rollbuhler, Phys. Rev. Letts. {\bf 94} 247205(2005).
\bibitem{degennes}P.G. De Gennes, Journal de Physique Lettres,{\bf 36} (3), 55-57(1975).
\bibitem{qcd} A. Ayalaab, A. Bashir, J.J.Cobos-Martínez, S Hernández-Ortiz and A. Rayac, Nuclear Physics B, {\bf 897} 77-86(2015)
\bibitem{lawrie} I.D. Lawrie  and S. Serbach,  ``Theory of tricritical points,'' Editors: C. Domb and J.L. Lebowitz, Vol. 9, Academic Press, New York, 1984.
\bibitem{barbosa} M.E. Fisher and M. C . Barbosa, {\bf 43}, 11177(1991);M.C. Barbosa and M.E. Fisher, Phys. Rev. B, {\bf 43} 10635(1991)
\bibitem{blume} M Blume, Phys. Rev. {\bf 141} 517(1966)
\bibitem{capel} H. W. Capel, Physica(Utrecht) {\bf 33},295(1967).
\bibitem{griffiths} R. B. Griffiths, Phys. Rev. Letts. {\bf 24} 715 (1970).
\bibitem{aizenman} M Aizenman and J Wehr,Phys. Rev. Lett., {\bf 62} 2503, (1989)
\bibitem{huiberker} K. Hui and A.N. Berker, Phys. Rev. Lett. {\bf 62},2507, (1989)
\bibitem{cardyj}J Cardy, Physica A {\bf 263} 215,1999; J. Cardy and J. L. Jacobsen, Phys. Rev. Lett. {\bf 79}, 4063(1997).
\bibitem{frazer} B.C. Frazer,G. Shirane,D.E. Cox and C.E. Olsen, J. Appl. Phys.,{bf 37} 1386 (1966). 
\bibitem{tfluids} D Mukamel and M Blume, Phys. Rev. A {\bf 10} 610 (1974).
\bibitem{salloys} K.E. Newman and J D Dow, Phys. Rev. B, {\bf 27} 7495(1983).
\bibitem{imelting} N Schupper and N.M. Shnerb, Phys. Rev. E {\bf 72} 046107(2005).
\bibitem{buzano} C. Buzano, A Maritan, A. Pelizzola, J. Phys. Cond. Matt. {\bf 6} 327(1994)
\bibitem{mfdilution} A Benyoussef,T Biaz,M Saber and M Touzani J. Phys. C {\bf 20} 5349(1987).
\bibitem{yuksel} Y Yuksel, U Akinci and H Polat, Physica A {\bf 391} 2819(2012).
\bibitem{branco} N.S. Branco and B.M. Boechat, Phys. Rev. B {\bf 56},11673(1997)
\bibitem{snowman} D. P. Snowman, Phys. Rev. E {\bf 79} 041126(2009)
\bibitem{bethe} E. Albayrak, Physica A {\bf 390} 1529(2011)
\bibitem{lara} D. P. Lara, Revista Mexicana de Fiscia {\bf 58} 203(2012).
\bibitem{salmon}  Octavio D Rodriguez Salmon and Justo Rojas Tapia, J. Phys. A:Math and General, {\bf 43}, 125003(2010).
\bibitem{santos} P.V. Santos, F.A de Costa and J.M. de Araujo, Physics Letters A{\bf 379} 1397(2015).
\bibitem{sumedha1} Sumedha, and Nabin Kumar Jana, J. Phys. A:Math and General, {\bf 50} 015003(2017).
\bibitem{touchette} H. Touchette, Physics Reports, {\bf 478}, 1(2009).
\bibitem{zirenberg} J. Zierenberg, N. G. Fytas, M Weigel, W. Janke and A. Malakis, European Physical Journal Special Topics, {\bf 226}, 789(2017)
\bibitem{silva} C. J. Silva, A. A. Caparica, and J. A. Plascak, Physical Review E, {\bf 73},036702(2006).
\bibitem{nelson} M. E. Fisher and D. R. Nelson, Phys. Rev. Lett. {\bf 32} 1350(1974).
\bibitem{helena-barbosa} E.L. de Santa Helena and M. C Barbosa, Physica A,{\bf 208},479(1994).
\bibitem{plascak} J. A. Plascak and D.P. Landau, Phys. Rev. E, {\bf 67} 015103(R), 2003.
\bibitem{butera} P. Butera and M. Pernici, Physica A, {\bf 507},22 (2018).
\bibitem{chaikin} P. M . Chaikin and T. M. Lubensky,"Principles of condensed matter physics", Cambridge Unbiversity Press, 1995.
\bibitem{upton}M. E. Fisher and P. J. Upton, Phys. Rev. Lett., {\bf 65} 2402(1990); Phys. Rev. Lett. {\bf 65} 3405(1990).
\bibitem{cardy} J. Cardy, "Scaling and renormalization in statistical physics", Cambridge Lecture Notes in Physics, (1996).
\bibitem{ellis} R.S. Ellis, Peter T Otto and H. Touchette, Annals of Applied Probability,{\bf 15},2203-2254(2005).
\bibitem{beg} M . Blume, V.J. Emery and R. B. Griffiths, Phys. Rev. A {\bf 4} 1071(1971).
\bibitem{mathematica} Wolfram Research, Inc. (n.d.). Mathematica
\bibitem{deepak} N. Vigneshwar, D. Mandal, K. Damle, D. Dhar and R. Rajesh, Phys. Rev. E, {\bf 99},052129(2019).
\bibitem{japan} Y. Ishibashi and Y. Hidaka, J. Phys. Soc. Japan, {\bf 60},11177(1991).
\bibitem{fytas} N.G. Fytas, J. Zierenberg, P.E. Theodorakis, M. Weigel, W. Janke, A. Malakis,Phys. Rev. E, {\bf 97} 040102(2019).
\bibitem{deserno} M. Deserno, Phys. Rev. E, {\bf 56} 5204(1997).
\bibitem{sethna} J. Kent- Dobias and J. P. Sethna, Phys. Rev. E, {\bf 98} 063306(2018).
\bibitem{rf} P. V. Santos, F. A da Costa and J. M de Araujo, J. of Magnetism and Magnetic Materials {\bf 451} 737(2018); M. Kaufman and M. Kanner, Phys. Rev. B, {\bf 42} 2378(1990).
\bibitem{wip} S. Mukherjee and Sumedha, work in progess.














\end{thebibliography}
\end{document}